# Computational Modelling of Atherosclerosis

Andrew Parton[1], Victoria McGilligan[1], Maurice O'Kane[2], Francina R Baldrick[1], Steven Watterson[1*]

[1]Northern Ireland Centre for Stratified Medicine, Ulster University, C-TRIC, Altnagelvin Hospital Campus, Derry, Co Londonderry, Northern Ireland, BT47 6SB
[2]Department of Clinical Chemistry, Altnagelvin Hospital, Western Health and Social Care Trust, Londonderry, Northern Ireland, BT47 6SB
*Corresponding author

Author description

Andrew Parton is a doctoral student at the Northern Ireland Centre for Stratified Medicine, Ulster University. His interests are in using computational approaches to stratify patients with cardiovascular disease.

Victoria McGilligan is an Assistant Professor at the Northern Ireland Centre for Stratified Medicine, Ulster University. Her interests are in inflammatory biomarkers for disease risk in personalized medicine.

Maurice O'Kane is a consultant chemical pathologist in the Western Health and Social Care Trust at Altnagelvin Hospital. His interests are in dyslipidaemia and clinical decision-making.

Francina R Baldrick is an Assistant Professor at the Northern Ireland Centre for Stratified Medicine, Ulster University. Her interests are in biomarkers for the role of nutrition in disease.

Steven Watterson is an Assistant Professor at the Northern Ireland Centre for Stratified Medicine, Ulster University. His interests are in computational modelling of cholesterol metabolism and cardiovascular disease.

Key Points

- Atherosclerosis is a disorder that emerges from a combination of dynamic processes, making it well suited to computational modelling.
- Atherosclerosis has been modelled to a range of levels of detail in recent work.
- There have been relatively few studies of plaque rupture and thrombosis with most work focussing on atheroma formation.
- Many elements of atherosclerosis have not yet been modelled which we describe.
- This is the first review to bring together the latest work in the area.

# Computational Modelling of Atherosclerosis

Andrew Parton[1], Victoria McGilligan[1], Maurice O'Kane[2], Francina R Baldrick[1], Steven Watterson[1*]

[1]Northern Ireland Centre for Stratified Medicine, Ulster University, C-TRIC, Altnagelvin Hospital Campus, Derry, Co Londonderry, Northern Ireland, BT47 6SB
[2]Department of Clinical Chemistry, Altnagelvin Hospital, Western Health and Social Care Trust, Londonderry, Northern Ireland, BT47 6SB
*Corresponding author

Atherosclerosis is one of the principle pathologies of cardiovascular disease with blood cholesterol a significant risk factor. The World Health Organisation estimates that approximately 2.5 million deaths occur annually due to the risk from elevated cholesterol with 39% of adults worldwide at future risk. Atherosclerosis emerges from the combination of many dynamical factors, including haemodynamics, endothelial damage, innate immunity and sterol biochemistry. Despite its significance to public health, the dynamics that drive atherosclerosis remain poorly understood. As a disease that depends on multiple factors operating on different length scales, the natural framework to apply to atherosclerosis is mathematical and computational modelling. A computational model provides an integrated description of the disease and serves as an *in silico* experimental system from which we can learn about the disease and develop therapeutic hypotheses. Although the work completed in this area to-date has been limited, there are clear signs that interest is growing and that a nascent field is establishing itself. This paper discusses the current state of modelling in this area, bringing together many recent results for the first time. We review the work that has been done, discuss its scope and highlight the gaps in our understanding that could yield future opportunities.

# Introduction

Cardiovascular disease (CVD) is the primary cause of death globally [1] and contributes to morbidity and mortality more than any other disorder in the western world [2]. In 2012, CVD was responsible for 31% of deaths worldwide, 47% of all deaths within Europe and 40% of all deaths within the European Union [3, 4]. CVD covers a collection of disorders that can be split into atherosclerotic and non-atherosclerotic categories [1]. Atherosclerotic CVD includes cerebrovascular disease [5], coronary artery disease [6] and peripheral vascular disease [7], and it is responsible for the majority of instances of CVD with a 2012 estimate attributing 71% of all CVD to atherosclerotic forms [4].

At least 75% of all CVD-related deaths occur in low and middle-income countries [3]. In China, more than 4% of the gross national income is directly spent on the treatment of CVD [8] and in the EU, it is estimated that CVD costs the economy approximately €196 billion per year [4]. Improvements in atherosclerosis and CVD treatment therefore have the potential to make a dramatic impact, not only on the quality of care available, but also on the economics of healthcare.

CVD is predominantly an age related condition. Coronary heart disease in men occurs five times more frequently in 80+ year old patients than similar patients in the 40-59 age group [9]. It is predicted that the global population aged 60+ will rise from 11% in the year 2000 [10] to 22% in 2050, making atherosclerosis a significant future public health concern. Comorbidities that drive CVD, such as diabetes [11], are set to grow with a global increase of 55% in cases projected between 2013 and 2035 [12]. The current and growing global risk of morbidity and mortality from atherosclerosis and the economic burden of treatment therefore make atherosclerosis an important area of future research.

Despite the growing importance of atherosclerosis and its implications for public health, its pathogenesis is not fully understood [13]. Traditionally, atherosclerosis was viewed as a build-up of lipids (including cholesterol) within the innermost layer of the artery wall (the *tunica intima*) [14].

However, our understanding has since developed and atherosclerotic CVD is now predominantly viewed as a chronic inflammatory condition, advanced by lipid build-up and triggered by innate immune responses [13, 15].

Atherosclerosis emerges as the results of multiple dynamical cell processes. Damage to endothelial cells [16] recruits monocytes to the site of inflammation via inter- and intra-cellular signalling [17]. Monocytes migrate into the artery wall [18], alongside lipoproteins, before differentiating into macrophages and phagocytosing oxidised low density lipoproteins (oxLDL) [19, 20]. The migration rate of these cells and particles is dependent upon haemodynamics [21] and vascular mechanical stress [22]. The accumulation of cholesterol-laden macrophages within the artery wall leads to plaque formation [23].

Studies aimed at understanding atherosclerosis need to be broad in scope and integrative in nature. The appropriate framework in which to consider emergent dynamical behaviour of this type is mathematical and computational modelling. A comprehensive programme of mathematical modelling and simulation can provide many benefits. Principally, it yields a framework for therapeutic hypothesis generation and for *in silico* drug target identification with the potential to streamline the drug development pipeline. This framework can be applied across populations or can be tuned to describe individual patients or patient groups as part of a programme of stratified, personalised and precision medicine [24].

Mathematical and computational models can take a range of forms. Ordinary differential equations (ODEs) [25], partial differential equations (PDEs) [25] and stochastic ordinary differential equations (SODEs) [26], alongside binary [27] and multivalued [28] logic have all been used to model pathway dynamics. Process algebras such as pi [29] and kappa [30] calculus have been used to capture the structure of pathway systems, in particular addressing the exponential growth in possible network configurations to be considered as the number of pathway components increases [31, 32].

Statistical models that infer pathway structure have been used to generate hypotheses from existing datasets [33, 34].

Computational biology approaches have previously been applied in studies of a range of dynamical disease processes. Examples include studies of Alzheimer's disease in which the pathways that mediate brain energy metabolism have been elucidated [35], studies of diabetes in which models of diagnostic testing have been developed along with models of the physiological mechanisms associated with disease [36] and studies of breast cancer in which biomarkers have been identified for the stratification of patient treatment [37]. Furthermore, computational models have been applied to pathway systems such as nuclear factor kappa beta (NF-κB) signalling [38], macrophage processing [39], human metabolism [40] and iron metabolism [41]. In one of the more ambitious computational studies of recent years, the first computational model of whole cell activity has appeared describing *Mycoplasma genitalium* [42].

Machine-readable standards for model representation have been developed to assist model development and model reuse. These standards have stimulated the creation of pathway informatics tools and have made models independent of the software tools used to create them. In particular, the Systems Biology Markup Language (SBML) [43, 44] and CellML [45] file formats capture ODE models describing the kinetics of pathway interactions and the Systems Biology Graphical Notation Markup Language (SBGN-ML) [46, 47] encodes diagrams of pathway function in a biologically meaningful file format. The Minimum Information Requested in the Annotation of Biochemical Models (MIRIAM) [48] and Minimum Information About a Simulation Experiment (MIASE) [49] standards describe model annotation and use respectively, and online repositories of SBML files have been introduced to facilitate model reuse [50].

Previously, cholesterol biosynthesis and the impact of therapeutic interventions have been modelled in a series of computational studies [51–55] and the role of lipid metabolism and CVD in aging has been reviewed [56]. However, no review has yet brought together the significant volume of recent work completed on computational modelling of atherosclerosis. This paper reviews the current state of this important nascent field, describing the work completed to date, discussing the approaches taken and highlighting the gaps in our understanding.

# The pathophysiology of atherosclerosis

In Figure 1, we see a representation of the processes that lead to atherosclerosis [13, 57, 58]. Damage to the endothelial layer of the artery wall triggers an inflammatory response in which monocytes, T-lymphocytes and other immune cells are recruited to the region of damage. These cells penetrate the endothelial layer, reaching the *tunica intima*, along with low density lipoprotein (LDL) and high-density lipoprotein (HDL) particles. Stimulated by the presence of interferon gamma (IFN-γ) and macrophage colony stimulating factor (M-CSF), monocytes differentiate into macrophages once they have entered the artery wall. While embedded within the *tunica intima*, both LDL and HDL become oxidized by free oxygen radicals. Macrophages will phagocytose oxidized LDL (oxLDL), but not oxidized HDL. Macrophages heavily loaded with oxLDL transform into foam cells that eventually undergo apoptosis. The resulting mass of debris embedded in the *tunica intima* is known as an atheroma. Foam cells, along with endothelial cells, secrete monocyte chemoattractant protein-1 (MCP-1) to recruit more monocytes to the site of inflammation. Naïve T cells contained within the artery wall differentiate into individual T cell types that can secrete IFN-γ. Smooth muscle cells (SMCs) are also recruited into the *tunica intima* where they undergo apoptosis and contribute to the formation of a fibrous cap in the artery wall. This accumulation of cells and debris can cause a swelling of the artery wall that restricts blood flow, leading to stenosis. If the fibrous cap ruptures, the build-up in the *tunica intima* is released into the blood stream increasing the risk of blockages downstream. Further complications can occur including clotting at the site of the atheroma where a thrombus forms further impeding blood flow.

# Computational modelling

## Blood flow dynamics

Vascular damage is a key trigger for the onset of atherosclerosis that can be induced by factors such as hypertension [59], smoking [60] and oxidative stress [61]. The elastic properties of arteries under hypertensive pressure have been modelled previously [62]. Obstructions to blood flow are known to be atherogenic [63] and it has been shown that this is in part attributable to the turbulent blood flow likely to be induced downstream [21, 22, 64–66].

A number of computational studies have modelled the dynamics of blood flow (*haemodynamics*) and its relationship to vascular structure. Navier-Stokes equations are typically used to describe blood flow through arterial structures [62–64, 66–82] under the assumption that blood flows as a Newtonian fluid, an approximation that can be violated by its viscosity and granularity. However, Navier-Stokes systems are well studied and are therefore represent a powerful framework for computational analysis. Arterial wall shear stress (WSS) is widely used as a model output that serves as a marker for atherosclerotic prone regions within an artery [21, 66, 67, 69–73, 76, 78–86]. How WSS impairs endothelial function is not well known, although its physiological significance has been demonstrated [22]. Two-dimensional and three-dimensional models of a Y-shaped arterial branch [66, 70, 71, 80, 83, 85–87] have been created along with linear artery models [64, 67, 68, 70–75, 78, 81, 88, 89]. Lower dimensional models are less physiologically accurate, but they provide the authors with more computationally amenable frameworks in which to demonstrate important principles, such as plaque stability [64] and the impact of stenosis [67].

Inflammation is thought to be driven by the penetration of the arterial wall by LDL, which in some cases is taken to be a function of the wall shear stress, demonstrating that an arterial branch can be a focal point for atheroma formation [66, 70, 71, 80, 81, 83, 85, 86]. As well as WSS, it has been shown that inflammation is related to blood viscosity, inlet flow rate and the geometry of the artery [73].

Simpler abstract models of this process have been developed that are less physiologically descriptive, but that enable more powerful mathematical approaches to be employed. They have been used to describe atherosclerosis as a bistable system for simple arterial geometries [84], to develop haemodynamic models in order to explore the turbulence downstream of an atherosclerotic constriction in two dimensions [88] and to describe haemodynamics and plaque formation as a test case for novel numerical methods [90].

## LDL concentration in the artery lumen

The turnover of LDL in the blood plays an important role as a primary factor that affects LDL penetration of the *tunica intima* in many models of atherosclerosis. Plasma LDL levels have been modelled as constant [71, 77, 80, 87], or as a variable [68, 70, 71, 74, 78, 81, 85, 90, 91] where the system dynamics are typically governed by a series of convection-diffusion equations, or part of a combined mass flow [82].

## LDL penetration of the tunica intima

The process through which LDL passes into the *tunica intima* has been modelled at a range of levels. The simplest approaches consider this to be a mathematical function of arterial WSS [92] or constant [77, 87, 90, 91, 93, 94]. Some simply ignore LDL penetration, instead considering only LDL in the *tunica intima* [84] or combining cells, proteins and other macromolecules into one mixed quantity [95]. More sophisticated approaches have considered diffusion [68, 74, 78, 82, 85, 96, 97] and have modelled the artery wall as a semi-permeable membrane by utilising Kedem-Katchalsky equations [67, 71, 81, 98, 99]. LDL penetration appears to be considered as a boundary to many models and the description of its uptake reflects the scope of the model proposed.

## LDL oxidation and the role of HDL

A range of approaches have been taken to describe LDL oxidation inside the *tunica intima* and they are coupled to LDL penetration to differing extents. Many studies consider the synthesis and turnover of oxLDL directly [68, 71, 80, 81, 84, 86, 87, 90, 91, 93, 94, 96–98, 100, 101]. In some, oxidation of LDL is a modelled reaction [68, 71, 80, 81, 86, 87, 90, 91, 93, 94, 96, 97] whereas in others it is taken to be a process that is driven by factors such as monocyte recruitment [84] or is modelled as a constant [100, 101]. Intermediate stages of the oxidation process have been considered by modelling the number of unoxidized antioxidant molecules attached to each LDL particle [94].

The role of HDL has been incorporated into a portion of these studies. In particular, it has been modelled as competing for free radicals and suppressing inflammatory signalling in the *tunica intima* [91, 94, 97] and as having an atheroprotective effect on foam cells [96, 101].

Elsewhere, the interplay between LDL, HDL, oxidising free radicals and antioxidant vitamins C and E have been studied [94] with predictions of comparable atheroprotective power between HDL and vitamin C.

## *Monocyte recruitment and chemoattractants*

Monocyte recruitment has been modelled as related to shear stress and the rate of LDL penetration [84]. The existence of monocytes in the lumen has rarely been considered [68], but several studies have modelled the turnover of monocytes in the *tunica intima* [68, 84, 91, 102, 103]. Elsewhere, the process of monocyte recruitment and differentiation has also been simplified and incorporated into one step governing macrophage turnover, where this is linked to driving factors such as shear stress, diffusion and LDL penetration [67, 80, 81, 86, 90, 93, 96–101].

Similarly, the turnover of MCP-1 as a chemoattractant has been described explicitly in some studies [96, 97] and grouped together with other chemoattractants including interleukin-1 (IL-1) and M-CSF in other studies [68, 86, 90, 91, 93, 99, 104]. One study has shown that exposure to radiation leads to enhanced levels of MCP-1 and is therefore atherogenic [103]. However, in many studies the role of chemoattractants has been ignored.

## *Monocyte to macrophage differentiation*

The differentiation of monocytes to macrophages has been incorporated into a number of studies, although many simplify this step by considering both populations as one group on the grounds that differentiation occurs on a time scale too short to be significant [67, 71, 100]. Where differentiation has been modelled it is presented with mass action kinetics [68, 84, 91, 103].

## Foam cell formation and the phagocytosis of oxidised LDL

The transformation of macrophages to foam cells due to the phagocytosis of oxLDL is a critical stage in the formation of atheroma that has been included in many studies. These are typically modelled as a combination of mass action and Michaelis-Menten terms [67, 68, 80, 81, 84, 86, 96, 97, 100, 101], and in some cases reverse cholesterol efflux is included in the model [96, 101, 105]. Many studies, however, omit foam cell formation as a step, instead taking the volume of macrophages recruited to be representative of atheroma formation [91, 93, 98, 102, 104].

## T cell recruitment and the role of interferon-gamma (IFN-γ)

The role of T cells in coordinating the inflammatory response has rarely been included in computational studies. Where they have been included as a factor, T cells yield IFN-γ that modulates macrophage differentiation [91, 96, 97, 103] and are themselves modelled as being activated by interleukin 12 (IL-12) [96, 97], although it has been shown experimentally that T cells can also be activated by IL-1 [106] and IFN-γ [107].

## Proliferation of smooth muscle cells

Along with foam cells and cell debris, SMCs contribute to the formation of atheroma [13]. However this factor has rarely been incorporated into models. Where it has been incorporated, SMC recruitment occurs in response to MCP-1, platelet derived growth factor (PDGF) and extracellular matrix (ECM) either modelled explicitly as factors [96, 97] or as a generic recruitment process [68, 90, 91, 93]. One study in particular has focused on the interplay between SMCs and PDGF identifying bistability in SMC-driven atheroma formation [108].

*Plaque rupture and thrombosis*

The rupture of atheroma has been modelled, establishing a criterion for atheroma instability that takes the form of a solution to a third order non-linear ODE [88]. Separate studies have established stability by evaluating the eigenvalues of a perturbed system [93] and by calculating the mean *time-to-rupture* of atheroma formation [102]. The WSS upon an atheroma has been calculated as a trigger for rupture and this model has been modified to incorporate the effects of abnormal axial G-forces [89]. Relevant models have been produced that describe thrombus formation in the absence [109] and presence of shear blood flow [64, 110].

# Discussion

The models described here are summarised in Table 1. The majority of the work presented has been published in the last 10 years, demonstrating that computational modelling of atherosclerosis is a developing field with growing support. These studies operate at a range of levels of abstraction and have variable scope. However, they have all been produced as separate bespoke computational models with little capacity for reuse by the wider modelling community. The adoption of community modelling standards such as SBML [43, 44] and SBGN-ML [46, 47] would enable the community to progress together on the development of atherosclerosis modelling and it would be a valuable exercise to translate the most biologically detailed models [68, 81] into these community standards.

Online databases, such as BioModels [50, 111, 112] and the Physiome Model Repository 2 (PMR2) [113], contain computational models of biological processes. Such databases facilitate the codification of our understanding and, critically, enable models to be reused and built upon as our knowledge advances. However, no models of atherosclerosis currently exist within these repositories, although systems biology representations of the cardiovascular system [114] and statin pharmacokinetics [115] are available.

The integration of models developed by different authors is likely to be a significant future challenge. The introduction of an online platform that presents networks of pathway diagrams using SBGN standards and enables users to select individual pathways or groups of pathways to study in isolation or to be downloaded for offline use, using SBGN-ML and SBML standards, would facilitate model reuse, maximising the value gained from their construction and enabling the research community to develop a coordinated consensus around the pathway biology more rapidly. Such a platform would not be limited to atherosclerosis but could be applied more broadly across pathway biology.

*Factors not yet modelled*

There are many components of atherosclerosis that to date have not been modelled. With accurate parameterisation each would increase the comprehensiveness and accuracy of our understanding of atherosclerosis as a dynamical process. Triglyceride rich lipoproteins contribute to plaque build-up with some studies showing that they trigger foam cell formation through mechanisms that bypass LDL oxidation [116–118]. Elsewhere, it has been proposed that categories of HDL and their relative proportions may be more important than the absolute abundance of HDL [57, 119], suggesting that models could be adapted to incorporate a HDL profile that influences oxidation and reverse cholesterol efflux. It has also been shown that HDL can inhibit the recruitment of monocytes and subsequently reduce atherogenesis [120] suggesting further interactions to model. Clinically, it has been suggested that LDL particle number is a stronger risk factor for atherosclerosis than the abundance of LDL-bound cholesterol, implying that future models should include a description of the cholesterol load of lipoproteins as well as their abundance [121]. In addition, the role of neutrophils [122], nitrous oxide [57], B cells [123], heat shock proteins [124, 125], sterol regulatory element binding protein (SREBP) mediated regulation [54], various cell signalling proteins such as NRLP3 [126], and miRNAs [127] have not been modelled in this context.

By far the majority of work to date has been on the build-up of atheroma. Some studies have addressed the mechanisms through which atheroma rupture, but they are in a significant minority. Very little work has been done on the consequences of rupture, such as thrombus formation. This presents a potential direction for the field that is highly relevant to patient treatment as most patients at risk of CVD are only identified after a cardiovascular event has occurred.

*Computational modelling in therapy development*

The application of computational modelling to therapy development in atherosclerosis has been historically poor. It is possible to predict both the efficacy of a drug and its potential side effects [128–130] and there is growing interest in areas of combinatorial drug design [131] to optimise treatment. Such approaches have been demonstrated for the role of statins in the reduction of LDL levels in plasma along with dietary changes [51, 97]. Computational biology can also be used to identify potential molecular targets for drugs and has been used to reduce the high attrition rate of drug discovery [133]. However, these technologies have yet to be exploited to their full potential. Finite element and analytical methods have been employed to model the interaction between a stent and artery wall when widening constricted arteries during angioplasty [134, 135].

Creating more comprehensive models of atherosclerosis has the potential to improve the efficiency of therapy development with benefits for both the patient and the commercial vendor. However, obtaining accurate parameterisations for the models is a fundamental challenge. The lack of appropriate published experimental data is a critical obstacle to generating high confidence predictive models.

*Difficulties in model generation*

Developing a comprehensive predictive model of atherogenesis comes with many challenges. Our knowledge of the processes involved has increased significantly in recent years with the development of genomic technologies such as genome wide association studies (GWAS) [136]. As atherosclerosis is a cardiovascular condition that affects critical circulatory systems, studying human atheroma poses logistical and ethical problems as access to live atherosclerotic tissue is limited and disturbances risk triggering plaque rupture. Consequently, data is limited. Animal studies of atherosclerosis do exist for mouse, rabbit and pig [137] and profiling has been conducted for plaque material removed in carotid endarterectomy [138].

The limited data available obstructs studies of atheroma at the macro scale and of the molecular biology involved. As a result, establishing biologically relevant kinetic parameters that can be used to simulate pathway dynamics is challenging and comprehensive parameterisations for the pathogenesis of atherosclerosis are not available in the public domain. As a consequence, some studies have resorted to estimating parameters for models, based on expert opinion or inferred from other cell processes.

It is likely that approximate values can be obtained for a number of the parameters required by using recombinant proteins and *in vitro* studies. However, recreating the environment of the *tunica intima* and quantifying its impact on the parameterisation in order to obtain physiological values will be challenging [139–141].

## Conclusion

Computational modelling of atherosclerosis presents an opportunity to contribute to the reduction of the global burden of CVD. By introducing accurate and quantitative models of atherosclerosis, we can create an *in silico* experimental system with the potential not only to displace *in vivo* and *in vitro* experimentation but also to enable us to study details that cannot be measured *in vivo* or *in vitro*. However this necessitates a physiologically accurate parameterisation and such data is not currently available in a comprehensive form.

Historically, little work has been completed developing computational modelling of atherosclerosis, although recent years have seen a clear growth of interest and the formation of a nascent field.

Here we have gathered together and reviewed the recent results with a view to identifying where the gaps in our understanding lie and where progress can be made.

Most of the work completed in this area to date has focussed on the inflammatory response and the shear stress of the artery wall and has involved modelling at a range of levels of abstraction.

The majority of work has focused on describing atheroma formation and few studies have addressed the mechanics of plaque rupture and its subsequent consequences. In most cases, models follow the canonical understanding of atherosclerosis: LDL penetration and oxidation, monocyte recruitment and differentiation and foam cell formation. However, many additional factors remain outside this canonical picture that are known to contribute to atherosclerosis and there currently exist opportunities to explore their role in the dynamics of this disease through computational modelling.

## Acknowledgements

We would like to acknowledge and thank Eliza Yankova (University of Ulster) for her assistance with Figure 1 (The pathophysiology of atherosclerosis).

## Table 1

**A summary of the mathematical models of atherosclerosis referenced within this review.**

These models are reproducible as their governing equations are explained in the cited references.

| First Author | Title | Year | Form | Parameters | Validation | Tools |
|---|---|---|---|---|---|---|
| Goriely [62] | On the mechanical stability of growing arteries | 2010 | A coronary artery modelled as an incompressible 2-layer cylindrical structure was used to study the arterial response to stress | Related to experimental data | Compared to experimental data obtained by Schulze-Bauer et al. [142] | None mentioned |
| Li [64] | How critical is fibrous cap thickness to carotid plaque stability? A flow-plaque interaction model | 2006 | A model of a stenotic carotid artery was used to relate fibrous cap thickness to WSS | Use a combination of estimated and experimentally validated parameters | The authors claim that the model fits well within the current literature, however no references are given to substantiate this claim | FEMLAB was used for model construction, SPSS was used to analyse this model |
| Stroud [66] | Numerical analysis of flow through a severly stenotic carotid artery bifurcation | 2002 | A model of a carotid artery bifurcation is used to study pulsatile and steady blood flow | Related to experimental data | Compared to experimental data obtained by Ahmed and Giddens [143] | None mentioned |
| Quarteroni [82] | Mathematical and numerical | 2002 | Proposed two models of an | Parameter source unclear | None | None mentioned |

| First Author | Title | Year | Form | Parameters | Validation | Tools |
|---|---|---|---|---|---|---|
| | modeling of solute dynamics in blood flow and arterial walls | | arterial bifurcation to study mass transfer | | | |
| Di Tomaso [67] | A Multiscale Model of Atherosclerotic Plaque Formation at Its Early Stage | 2011 | Built a multi-scale model of atherosclerosis to include mass transfer, LDL oxidation and foam cell formation | Use a combination of estimated and experimentally validated parameters | The model was compared with experimental data taken from Meyer *et al.* [144] and against the model produced by Olgac *et al.* [74] | None mentioned |
| Cilla [68] | Mathematical modelling of atheroma plaque formation and development in coronary arteries | 2013 | Uses a standard left descending coronary artery model to study plaque growth | Taken from experimental data and other mathematical models | Parts of the model correspond with experimental data such as Meyer *et al.* [144], however appropriate experimental data to cover the entire model is not currently available. | COMSOL Multiphysics |
| Filipovic [70] | Computer simulations of blood flow with mass transport through the carotid artery bifurcation | 2004 | Proposed a simulation of mass transport to allow physicians to study individual patients | Parameter source unclear | The authors claim that the model fits well within the current literature, however no references are given to substantiate this claim | None mentioned |

| First Author | Title | Year | Form | Parameters | Validation | Tools |
| --- | --- | --- | --- | --- | --- | --- |
| Filipovic [71] | ARTreat Project: Three-Dimensional Numerical Simulation of Plaque Formation and Development in the Arteries | 2012 | Presented a 3D model of plaque formation and development | Parameters were experimentally established through a rabbit animal model | Plaque progression within the model has been validated against experimental data taken from Boussel *et al.* [145] | None mentioned |
| Johnston [72] | Non-Newtonian blood flow in human right coronary arteries: Transient simulations | 2005 | Used right coronary artery models to study pulsatile blood flow | Experimentally observed | Findings were validated against experimental data taken from Kirpalani *et al.* [146] & Myers *et al.* [147] | CFD-ACE |
| Liu [73] | Computer Simulations of Atherosclerotic Plaque Growth in Coronary Arteries | 2010 | Uses model of a stenosis-free curved human coronary artery to study plaque growth | Experimentally observed | None | COMSOL Multiphysics |
| Olgac [74] | Computational modeling of coupled blood-wall mass transport of LDL: effects of local wall shear stress | 2008 | Developed a model of a stenosed coronary artery to study the effects of WSS on mass transport | Experimentally observed | Related to experimental data Meyer *et al.* [144], Huang *et al.* [148], Yuan *et al.* [149] | COMSOL Multiphysics |
| Rappitsch [75] | Numerical Modelling of Shear-Dependent Mass Transfer in Large Arteries | 1997 | Used a curved-tube-artery model to study blood flow and lipoprotein transport processes | Use a combination of estimated and experimentally validated parameters | Validated against Friedman *et al.* [150] | None mentioned |

| First Author | Title | Year | Form | Parameters | Validation | Tools |
| --- | --- | --- | --- | --- | --- | --- |
| Sun [76] | Fluid-wall modelling of mass transfer in an axisymmetric Stenosis: Effects of shear-dependent transport properties | 2006 | Studies the influence of WSS on mass transport | Use a combination of estimated and experimentally validated parameters | Compared to experimental data taken from rabbit aortic walls Meyer *et al.* [144] | None mentioned |
| Ai [77] | A coupling model for macromolecule transport in a stenosed arterial wall | 2006 | A model of a stenosed artery is used to study lipid transfer | Experimentally validated | Compared to other mathematical models, with arguments as to why their parameter set is more accurate | FIDAP |
| Wada [78] | Theoretical study of the effect of local flow disturbances on the concentration of low-density lipoproteins at the luminal surface of end-to-end anastomosed vessels. | 2002 | Femoral artery model is used to study the relationship between intimal thickness and the endothelial surface level of LDL | Parameters were taken from experimental data or estimated | Compared to experimental data taken from Ishibashi *et al.* [151] | Star LT |
| Calvez [80] | Mathematical modelling of the atherosclerotic | 2009 | Developed a 2D geometry modelling the | Parameters are taken from other mathematical | None | FreeFem++ [153] |

| First Author | Title | Year | Form | Parameters | Validation | Tools |
|---|---|---|---|---|---|---|
| | plaque formation. | | carotid artery to demonstrate plaque formation, based on the model of El Khatib *et al.* [104] | models, relating to atherosclerosis [104] and hyperplasia [152] | | |
| Calvez [81] | Mathematical and numerical modelling of early atherosclerotic lesions | 2010 | Expanded on their previous model [80] to include a model of lesion growth | Parameters were taken from experimental data or estimated | Experiments published by Cheng *et al.* [154, 155] were reproduced and were used to validate the model | FreeFem++ [153] |
| Bosnić [83] | Mining data from hemodynamic simulations for generating prediction and explanation models. | 2012 | Built a prototype of a system that could predict locations of increased WSS from artery models | Parameter source unclear | Presents a series of methods to estimate accuracy of the model, and relates these to experimental data | None mentioned |
| Bulelzai [84] | Long time evolution of atherosclerotic plaques | 2011 | Present a series of ODEs for the concentrations of particular elements of atheromae. | Taken from experimental data and other mathematical models [91] | Compared to mathematical model of Zohdi *et al.* [102] | MATCONT [156] |
| Gabriel [85] | Deposition-driven Growth in Atherosclerosis Modelling. | 2014 | A simplified bifurcating artery is used to model LDL flux into the intima | Taken from experimental data | None | ANSYS Fluent |
| Silva [86] | Mathematical Modeling of Atherosclerotic | 2013 | Built a 2D carotid artery bifurcation to study plaque | Taken from other mathematical models [157] | None | COMSOL Multiphysics |

| First Author | Title | Year | Form | Parameters | Validation | Tools |
|---|---|---|---|---|---|---|
| | Plaque Formation Coupled with a Non-Newtonian Model of blood Flow | | formation with a non-Newtonian model of blood flow | | | |
| Gessaghi [87] | Growth model for cholesterol accumulation in the wall of a simplified 3D geometry of the carotid bifurcation | 2011 | A 3D model of a carotid artery bifurcation is used to study the influx, efflux, oxidation and phagocytosis of LDL | Taken from experimental data | Compared with data obtained from Yang et al.[158]. However, authors comment that not enough experimental data exists to fully validate the model. | OpenFOAM [159], Netgen |
| Green [88] | Atherosclerotic plaque rupture | 2002 | A model of a straight, stenotic 2D artery is used to study atherosclerotic plaque rupture. | Parameter source unclear | None | AUTO |
| Deepa [89] | Modelling Blood Flow and Analysis of Atherosclerotic Plaque Rupture under G-Force | 2009 | A 1D arterial model was used to study the rupture of plaques under g-force | Sources have not been cited for parameter values | None | MATLAB |
| Girke [90] | Efficient Parallel Simulation of Atherosclerotic Plaque Formation Using Higher Order Discontinuous Galerkin Schemes | 2014 | Girke et al. built a mathematical model based on the works of Ibragimov et al. [93] and Calvez et al. [81] to | Taken from experimental data | None | DUNE-FEM [160] |

| First Author | Title | Year | Form | Parameters | Validation | Tools |
|---|---|---|---|---|---|---|
| | | | demonstrate the use of the compact discontinuous galerkin method (CDG2) in discretizing relevant equations | | | |
| McKay [91] | Towards a Model of Atherosclerosis | 2005 | Proposed a mathematical model to cover mass transfer, oxidation, immune cell activation and plaque growth | Taken from other mathematical models, or estimated by domain experts | None | None mentioned |
| Ibragimov [93] | A mathematical model of atherogenesis as an inflammatory response | 2005 | Created a series of ODEs to study the concentrations of cell groups over time | Primarily estimated due to lack of relevant data | None | FEMLAB |
| Cobbold [94] | Lipoprotein Oxidation and its Significance for Atherosclerosis: a Mathematical Approach | 2002 | Built a series of ODEs to study lipoprotein oxidation | Taken from experimental data | Compared to an experiment performed by Neužil *et al.* [161] | None mentioned |
| Prosi [95] | Mathematical and numerical models for transfer of low density lipoproteins through the arterial walls: a new | 2004 | Built multiple models of lipoprotein transfer in order to maximse the accuracy of their prediction | Taken from experimental data | Experimentally validated against Meyer *et al.* [144] | None mentioned |

| First Author | Title | Year | Form | Parameters | Validation | Tools |
|---|---|---|---|---|---|---|
| | methodology for the model set up with applications to the study of disturbed lumenal flow | | | | | |
| Friedman [96] | A Mathematical Model of Atherosclerosis with Reverse Cholesterol Transport and Associated Risk Factors | 2014 | Expands on the previous model by the same group [97] to include reverse cholesterol transport | Taken from experimental data and from other mathematical models [97] | Validated qualitatively against experimental data (e.g. [162–164]) | None mentioned |
| Hao [97] | The LDL-HDL profile determines the risk of atherosclerosis: a mathematical model | 2014 | Developed a series of PDEs to model the concentration of a series of cells and macromolecules contained within an atheroma, and related this information to plaque growth | Taken from experimental data or estimated | None | MATLAB |
| Filipovic [99] | Experimental and computer model of plaque formation in the artery | 2011 | Built a model of plaque formation based on a pig left anterior descending coronary artery | Taken from experimental data, or estimated where data was unavailable. | Reproduced an experiment by Cheng et al. [155] and compared the results to their model of plaque formation | None explained |

| First Author | Title | Year | Form | Parameters | Validation | Tools |
|---|---|---|---|---|---|---|
| Ougrinovskaia [100] | An ODE model of early stages of atherosclerosis: mechanisms of the inflammatory response | 2010 | Developed a series of ODEs to model mass transfer and foam cell formation | Estimated | Behaviour relates to qualitative data, but model has not been compared to quantitative data | MATLAB, XPPAUTO |
| Cohen [101] | Athero-protective effects of High Density Lipoproteins (HDL): An ODE model of the early stages of atherosclerosis | 2014 | Expanded on their previous model [100] to include HDL and reverse cholesterol transport | Taken from experimental data | Noted that the behaviour of their model corresponds with an experiment performed by Feig *et al.* [163] | None mentioned |
| Zohdi [102] | A phenomenological model for atherosclerotic plaque growth and rupture | 2004 | Built a series of equations to study plaque growth and lesion rupture | Taken from experimental data, or estimated where data was unavailable. | None | None mentioned |
| Little [103] | A model of cardiovascular disease giving a plausible mechanism for the effect of fractionated low-dose ionizing radiation exposure | 2009 | Built a series of equations to study the effect of small radiation doses to atherosclerosis and CVD | Taken from experimental data | Sections of this model are validated by matching with experimental data published by Cushing *et al.* [165] and Shi *et al.* [166] | None mentioned |
| El Khatib [104] | Atherosclerosis Initiation Modeled as an Inflammatory Process | 2007 | Built a series of reaction-diffusion equations by grouping together | Estimated | None | COMSOL Multiphysics |

| First Author | Title | Year | Form | Parameters | Validation | Tools |
|---|---|---|---|---|---|---|
| | | | all cytokines and immune cells involved | | | |
| Zhang [105] | Foam cell formation in atherosclerosis: HDL and macrophage reverse cholesterol transport | 2013 | Expanded on the model of Ibragimov *et al.* [93] by focusing on the role of HDL and reverse cholesterol transport | Taken from experimental data and from other mathematical models [94] | None | None mentioned |
| Fok [108] | Mathematical model of intimal thickening in atherosclerosis: vessel stenosis as a free boundary problem | 2012 | Focuses on SMC migration and the role of PDGF | Taken from experimental data | Compared to experimental data taken from New Zealand white rabbits Stadius *et al.* [167] | None mentioned |
| Xue [109] | A mathematical model of ischemic cutaneous wounds | 2009 | Xue *et al.* developed a series of PDEs to model ischemic dermal wounds | A combination of experimentally validated and estimated parameters are used | Compared to experimental data established by Roy *et al.* [168] | Livermore Solver |
| Guy [110] | Fibrin gel formation in a shear flow | 2007 | Presents a model of fibrin formation in a damaged blood vessel | A combination of experimentally validated and estimated parameters are used | None | None mentioned |

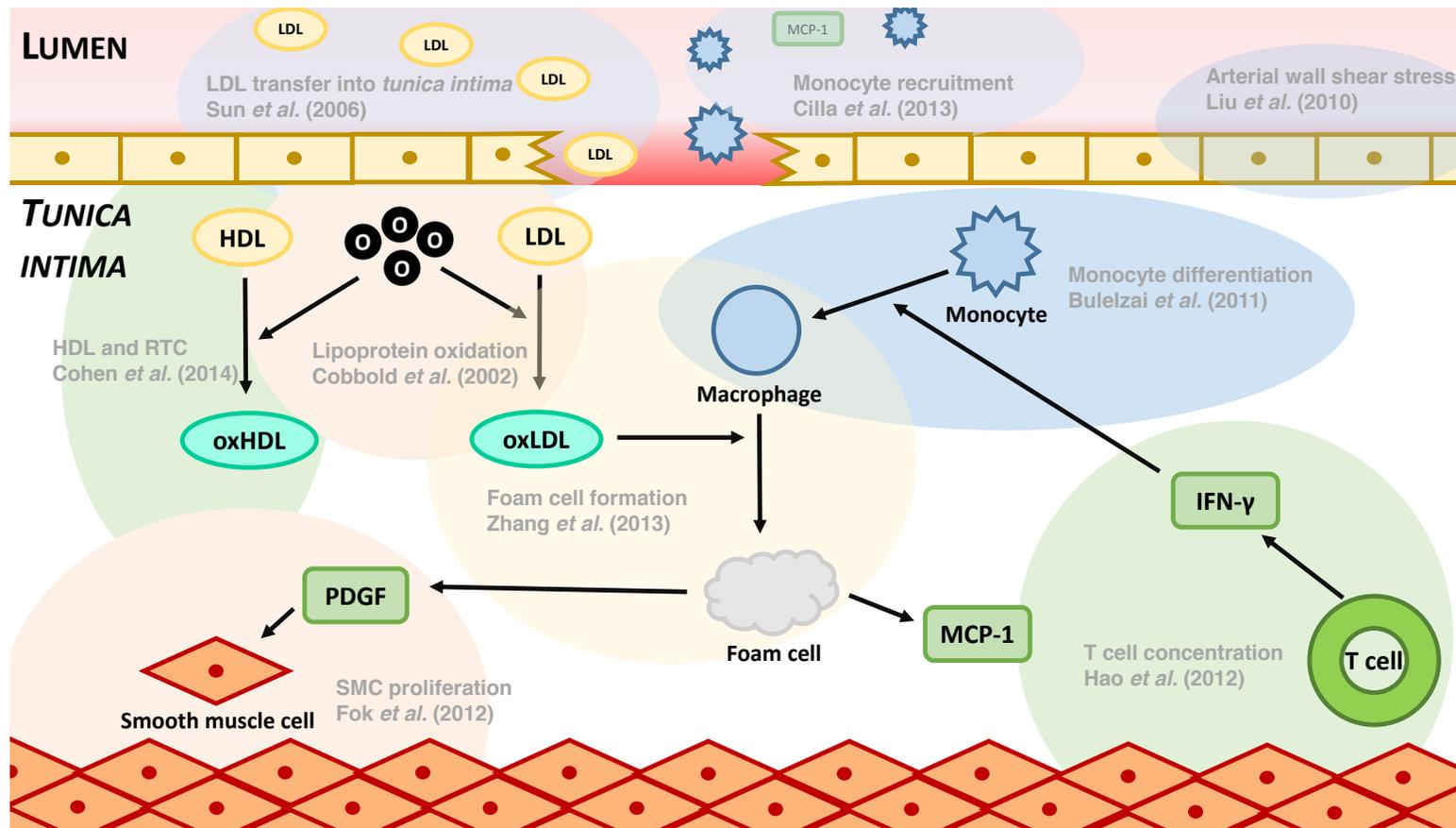

**Figure 1: The pathophysiology of atherosclerosis.** Low density lipoproteins (LDL) transfer into the artery wall at a site of endothelial damage. Arterial wall shear stress (WSS) and its relationship to lipoprotein transfer into the artery wall has been studied by Liu *et al*. Lipoproteins pass into the artery wall at a rate dependent on WSS, lipoprotein diffusivity and lipoprotein concentration, as modeled by Sun *et al*. After entering the intima, lipoproteins become oxidized upon contact with free oxygen radicals, a process that has been modeled by Cobbold *et al*. Monocytes are recruited to the site of inflammation via MCP-1 (modeled by Cilla *et al.*) and pass into the intima before differentiating into macrophages (Bulelzai *et al.*), catalyzed by T-Cell produced IFN-γ (Hao *et al.*). Macrophages phagocytose oxidized LDL within the artery wall, forming cholesterol-laden foam cells (Zhang *et al.*). Foam cells secrete MCP-1, which recruits more monocytes to the lesion, and PDGF, which recruits smooth muscle cells (SMCs) into the intima (Fok *et al.*).